\documentstyle[preprint,prl,aps]{revtex}

\begin{document}
\draft
\preprint{}
\title{Vanishing spectral weight in a one-hole-doped antiferromagnet:\\
A rigorous result }
\author{Z. Y. Weng, Y. C. Chen,$^*$ and D. N. Sheng}
\address{Texas Center for Superconductivity and Department of Physics\\
University of Houston, Houston, TX 77204-5506 }
\maketitle
\begin{abstract}

By explicitly tracking  the Marshall sign, a phase string induced by hopping is
revealed in the one-hole-doped $t-J$ model. It is rigorously shown that
such a phase string cannot be eliminated through low-lying spin excitations,
and it either causes the spectral weight $Z=0$, or leads to
the localization of the doped hole. Implications for finite doping are also
discussed.
\end{abstract}

\vspace{0.4in}

\narrowtext

Whether the strongly-correlated $t-J$ model can be meaningfully
treated in terms of a conventional perturbative method is still controversial.
The key issue\cite{anderson} involves  the spectral weight $Z$. If $Z=0$ in an
interested regime, it means that each particle (or hole) added to the system
cannot be described as a quasi-particle-type excitation. Instead, each
injected particle would cause a global change in the original ground state,
which may not be tractable perturbatively. In particular, the issue whether
$Z=0$ for a one-hole doped anitiferromagnet is crucial for
establishing a workable method at weak doping: a finite $Z$ would mean that
at least in dilute doping the low-lying state may be described by some
quasiparticle-type holes on the top of an antiferromagnetic (AF) background,
with {\it weak} interaction among themselves. However, $Z=0$ would simply
suggest
the breakdown of this perturbative connection between the weakly-doped system
and the parental antiferromagnet.

The discussions about the spectral weight $Z$ for a one-hole problem in
various analytical approaches\cite{anderson,kane} and finite-size numerical
calculations\cite{dagotto} have given
conflicting conclusions in the two-dimensional (2D) case. The main difficulty
involved in this problem
arises from the competing superexchange and hopping processes, which usually
result in a strong spin-polaron distortion around the doped hole. Such a spin
distortion effect constitutes a dominant part in the energy spectrum of the
renormalized hole. Most importantly, the spin-polaron is different from the
usual
phonon-polaron as $SU(2)$ spins are involved in the former, where $U(1)$
phase may play an important role in shaping the long-distance part of
spin-polaron with {\it little} energy cost.  In a
finite-size calculation and truncated self-consistent approximation, there is
no accurate way to distinguish the contributions from the coherent bare-hole
part and  the rest  ``spin-polaron'' part in  a
sufficiently low-energy and long-wavelength regime. Therefore,
a more accurate approach would be desirable in order to get access to such a
regime. In this Letter, we introduce a spin-hole basis in which the Marshall
sign as the sole source of sign problem hidden in the spin background can be
exactly tracked. A new
phase string induced by the doped hole is then revealed in this representation.
Due to such a phase string, one either has $Z=0$, or the doped hole is
localized in the  ground state of the one-hole doped $t-J$ model.

Let us start with the undoped antiferromagnet. It is described by the
superexchange Hamiltonian
\begin{equation}\label{e01}
H_J= J\sum_{<ij>} \left[{\bf S}_i\cdot {\bf S}_j -\frac {n_in_j}{4}\right],
\end{equation}
which is equivalent to the Heisenberg model at half-filling as the occupation
number $n_i=n_j=1$. According to Marshall\cite{marshall}, the ground-state
wavefunction of the Heisenberg Hamiltonian for a bipartite lattice is real and
satisfies a sign rule. This sign rule can be uniquely determined by requiring
that a flip of two antiparallel spins at nearest-neighboring
sites always involves a sign change: $\uparrow\downarrow \rightarrow (-1)
\downarrow\uparrow$ in the wave-function. The meaning of the Marshall sign may
be understood as follows: under
a spin basis $|\phi>$ with the Marshall sign included, the matrix element of
the superexchange Hamiltonian (\ref{e01}) becomes negative-definite, i.e.,
$<\phi'|H_J|\phi>\leq 0$.
Then it is easy to see that the ground-state $|\psi_0>=\sum_{\phi}c^0_{\phi}|
\phi>$ always
has real, positive coefficient (or wave-function) $c^0_{\phi}$, so that the
aforementioned Marshall sign is indeed the only sign to appear in the ground
state $|\psi_0>$ (through $|\phi>$). And there is no additional sign problem
in this new representation.

The Marshall sign described above can be easily built into the $S^z$-spin
representation even in the presence of a hole. The bipartite lattice can be
divided into even ($A$)
and odd ($B$) sublattices. For each down spin at $A$ site or up spin at $B$
site, one may assign an extra phase $i$ to the basis. In this way, a flip of
two nearest-neighboring spins will always involve a sign change (i.e., the
Marshall sign): $\uparrow\downarrow \rightarrow (i^2)\downarrow\uparrow
= (-1)
\downarrow\uparrow$. Of course, this is not a unique way to incorporate the
Marshall sign in the spin basis, but it will be quite a useful bookkeeping
once hole is introduced into the system. Generally speaking, the spin
basis with one hole may be defined as
\begin{equation}\label{e1}
|\phi_{(n)}>= i^{N_A^{\downarrow}+N_B^{\uparrow}} |\uparrow
...\downarrow\uparrow
\circ ...\downarrow>  ,
\end{equation}
with $N_A^{\downarrow}$ and $N_B^{\uparrow}$ as total numbers of
down and up spins
at A- and B-sublattices, respectively, and the subscript $(n)$ specifying the
hole's site. It is straightforward to verify that the matrix element
\begin{equation}\label{e2}
<\phi'_{(n)}|H_J|\phi_{(n)}>\leq 0,
\end{equation}
under this new basis. Namely, even in the presence of a hole, the superexchange
term $H_J$ still does not have sign problem in the representation (\ref{e1}).

Now we consider the hopping of the hole. The hopping process is governed by
$H_t$ defined by
\begin{equation}\label{ehopping}
H_t= -t \sum_{<ij>} c_{i\sigma}^{\dagger}c_{j\sigma} + H.c.  ,
\end{equation}
where the Hilbert space is restricted by the no-double-occupancy constraint
$\sum_{\sigma}c_{i\sigma}^{\dagger}c_{i\sigma}\leq 1$. Suppose the hole
initially sitting at site $n$ hops onto a nearest-neighboring site $m$. The
corresponding matrix element in the basis (\ref{e1}) can be easily obtained:
\begin{equation}\label{e3}
<\phi_{(m)}|H_t|\phi_{(n)}>= -t \times (\pm i) ,
\end{equation}
where the sign $\pm $ is determined by the original site-$m$ spin state
$\sigma_m=\pm 1$ in the following way:
\begin{equation}\label{e4}
(\pm i) \equiv  (-1)^m e^{i\frac {\pi} 2 \sigma_m}.
\end{equation}
Here $(-1)^m$ is the staggered factor: $(-1)^A=+1$ and $(-1)^B= -1$.
Thus, in terms of the representation (\ref{e1}), it is revealed that the hole
will always leave behind a trace of phases (phase string) $(\pm i)\times (\pm
i)\times ...$ when it moves around. On the other hand, if there
exists a long-range AF ordering in the ground state, the hopping
process in (\ref{e3}) also creates a well-recognized ``string'' defect\cite
{dagotto,trugman},
namely, a line of mismatched spins along the polarization direction left by the
moving hole. However, this latter type of string
can be dynamically eliminated through spin-flip process described in
(\ref{e2}). Instead of causing the localization of the hole,
it merely leads to  a reduction of  the spectral weight $Z$ by $J/t$\cite{kane}
(or to some power of $J/t$\cite{kane,dagotto}).  In contrast, the phase string
induced by hopping (\ref{e3}) cannot be repaired through the spin process
(\ref{e2}) at low energy (see below). Therefore, one expects
such a phase string to be essential in determining the long-wavelength
behavior of the hole.   This phase problem is closely related to a $U(1) $
subgroup of the $SU(2)$ spins, whose importance has been already revealed in
the
one dimensional (1D) case\cite{weng1}.  In the following,  we will investigate
its crucial role for a general dimensionality.

We define a ``bare''  hole to be described by $c_{i\sigma}|\psi_0>$.
One can track its evolution by studying the propagator $G_{1\sigma}(i,j;
E)=<\psi_0|c_{j\sigma}^{\dagger}
(E-H_{t-J}+i\eta)^{-1}c_{i\sigma}|\psi_0>$, with $H_{t-J}=H_t+H_J$ and
$\eta=0^+$.
In momentum space, the imaginary part of $G_{1\sigma}(k, E)$ is given by
\begin{equation}\label{eim}
\mbox{Im} G_{1\sigma}(k, E)=-\pi \sum_M Z_k(E_M) \delta(E-E_M),
\end{equation}
where the spectral weight $Z_k$ is defined as
\begin{equation}
Z_k(E_M)=|<\psi_M|c_{k\sigma}|\psi_0>|^2,
\end{equation}
with $|\psi_M>$ and $E_M$ denoting the eigenstate and energy of $H_{t-J}$ in
the one hole case. The corresponding real-space form of (\ref{eim}) may be
rewritten as
\begin{equation}\label{eg"}
G_{1\sigma}''(j,i;E)=-\pi \sum_k e^{-ik\cdot (x_j-x_i)}Z_k(E) \rho(E),
\end{equation}
where $\rho(E)=\sum_M\delta(E-E_M)$ is the density of states, and here
$Z_k$ is understood as being averaged over $M$ at the same energy $E$.

Spectral weight $Z_k(E)$ describes the overlap of the bare-hole state $c_{k
\sigma}|\psi_0>$ with the real eigenstates at energy $E$. If the low-lying
excitation can be classified as quasiparticle-like, $Z_k(E)$ must be
finite and peaked at a lower-bound energy $E_k$ near the ground-state
energy $E_G$,  with a well-defined dispersion
relation with regard to the momentum $k$. Particularly, at the band bottom
$Z_k(E_G)$ should be finite at some {\it discrete} $k$'s  determined by
$E_k=E_G$.
In terms of  (\ref{eg"}), then, $G_{1\sigma}$ must  become sufficiently
extended in a large scale $|x_j-x_i|$ when $E$ is close enough to $E_G$.   In
other words, if one finds $G_{1\sigma}\rightarrow 0$ as $|x_j-x_i|\rightarrow
\infty$ even
near $E= E_G$, $Z_k(E_G)$ has to be either zero\cite{remark} or  involves a
continuum of $k$'s to lead a decay in (\ref{eg"}).  In the latter case,  $E_k$
is dispersionless near the band bottom such that  the hole in the ground state
has an infinite effective mass and is localized in space.

To separate the hopping and superexchange processes,
one may expand $G_{1\sigma}$ in terms of $H_t$ as follows
\begin{equation}
G_{1\sigma}(j,i;E)=<\psi_0|c_{j\sigma}^{\dagger}\left(
G_0^J+G_0^JH_tG_0^J+... 	\right)
c_{i\sigma}|\psi_0>,
\end{equation}
where $G^J_0=(E-H_J+i\eta)^{-1}$. Then, by inserting intermediate
states in terms of the representation of (\ref{e1}),  one
obtains
\begin{eqnarray} \label{epath}
G_{1\sigma}(j,i; & E)&=e^{i\frac {\pi} 4 \sigma [(-1)^j-(-
1)^i] }\sum_{\mbox{(all paths)}}\sum_{\mbox{(all  states)}} c^0_{\phi'_{(j)}}
c^0_{\phi_{(i)}} \nonumber\\ & & \times T_{ij}(\{\phi\})
\prod_{s=0}^{K_{ij}}<\phi_{(m_s)}^{s+1}|G_0^J(E)|\phi_{(m_s)}^s>,
\end{eqnarray}
which involves those intermediate states $\{|\phi_{(m_s)}^{s, s+1}>\}$ where
the hole is on a path
$m_0, m_1, ..., m_{K_{ij}} $ connecting sites $i$ and $j$. Here $m_0=i$,
$m_{K_{ij}}=j$, and $K_{ij}$ denotes the number of
links for a given path.  And $|\phi_{(m_0)}^0>\equiv |\phi_{(i)}>$, $|\phi_{(
m_{K_{ij}})}^{K_{ij}+1}>\equiv |\phi_{(j)}'>$. $T_{ij}$ is the
contribution from $H_t$ for such a path:
\begin{equation}\label{estring}
T_{ij}(\{\phi\})=(-it)^{K_{ij}}\prod_{k=1}^{K_{ij}}(-1)^{m_k}\sigma_{m_k},
\end{equation}
with $\sigma_{m_k}$ specifies the site-$m_k$ spin state in the state $|\phi_{
m_{k-1}}
^{k}>$ where the hole is at site $m_{k-1}$.
Equation (\ref{estring}) shows that each path connecting sites $i$ and $j$ is
weighted by
a spin-dependent phase string $\prod_{k=1}^{K_{ij}}(-1)^{m_k}\sigma_{m_k}$.
The rest phases {\it explicitly} shown in (\ref{epath}) and (\ref{estring}) are
constant for each
path,  and thus are trivial.  (In fact, those phases lead to oscillations of
the propagator corresponding to momenta, e.g.,($\pm \pi/2$, $\pm \pi/2$)  in
2D, which is well-known in a single-hole doped problem\cite{kane}.)

To determine the phase of
$<\phi'_{(m)}|G_0^J|\phi_{(m)}>$ in (11), one may expand $G_0^J$ as follows
\begin{equation}
G_0^J(E)=\frac 1 E \sum_k \frac {{H_J}^k}{E^k}.
\end{equation}
By using $<\phi_{(m_s)}^{s+1}|{H_J}^k|\phi_{(m_s)}^s>= (-1)^k
|<\phi_{(m_s)}^{s+1}|{H_J}^k|\phi_{(m_s)}^s>| $ according to (\ref{e2}) ,
one finds
\begin{equation}\label{eg0}
<\phi_{(m_s)}^{s+1}|G_0^J(E)|\phi_{(m_s)}^s>=\frac 1 E \sum_k \frac {
|<\phi_{(m_s)}^{s+1}|{H_J}^k|\phi_{(m_s)}^s>|}{(-E)^k} <0 ,
\end{equation}
if $E<0$. It is easy to see that $E<E_G^0$ (the lowest
energy bound  of $H_J$  with the hole localized
at site $m_s$) will  guarantee the convergence of  (\ref{eg0}). So the matrix
element $<\phi'_{(m)}|G_0^J|\phi_{(m)}>$ appearing in (\ref{epath}) is always
sign-definite at low energy.
Therefore, the phase string induced by the doped hole in (\ref{estring}) indeed
is not  ``repairable'' by the low-lying spin fluctuations in the energy
regime $E_G^0>E\geq E_G$. This is consistent with the previous intuitive
observation, and implies important consequences for long-wavelength
behaviors of a doped hole.

Now we focus on the contribution of the phase string $\tilde{T}_{ij}\equiv
\prod_{k=1}^{K_{ij}}(-1)^{m_k}\sigma_{m_k}$ in
(\ref{epath}).  It is convenient for one to define
\begin{equation}
\tilde{\sigma}_m=(-1)^k\sigma_m.
\end{equation}
The physical meaning of  $\tilde{\sigma}_m$ is as follows: if there is a
fully-polarized N\'{e}el order in the spin background, one has $\tilde{
\sigma}_m\equiv 1$ (or $-1$ by
shifting the ordering by one lattice constant).  Then $\tilde{\sigma}_m= -
1$ represents a ``spin-flip'' at  site $m$ with regard to such a
N\'{e}el configuration. This N\'{e}el configuration is just a
reference for describing various intermediate spin configurations in
(\ref{epath}), and
the phase string is contributed by all those ``spin flips'' in terms of
$\prod_{k=1}^{K_{ij}}\tilde{\sigma}_{m_k}=
(-1)^{N^{flip}_{ij} } $, where $N^{flip}_{ij} $ is the
total number of  zero-point ``spin flips''  on the given path connecting $i$
and $j$.  Thus the effect of the zero-point spin fluctuations is
{\it accumulated} in the phase string.  In other words,  with the increase of
length scale,  more and more spin-flips will be involved, with each of them
contributing a $(-1)$.  Such an effect, then, must lead to a vanishing
contribution to the propagator (\ref{epath}) when $|i-j|\rightarrow \infty$.
The proof is quite straightforward.   One may define the contribution of
the phase string in (\ref{epath}) as $<e^{i\pi N_{ij}^{flip}}>_{path}$, where
$<...>_{path}$ means the average over all the spin states for a given path,
weighted
by  other sign-definite quantities shown in (\ref{epath}).  If  $<e^{i\pi
N_{ij}^{flip}}>_{path}$ would approach to a finite quantity $B_{\infty}<1$
(omitting some oscillation factor) beyond a large scale $L$ ($L\gg$ the
spin-fluctuational correlation length), then at a scale
$\sim r\times L$ ($r>1$),  one would have $<e^{i\pi N_{ij}^{flip}}>_{path}
\simeq (B_{\infty})^r$ to the leading order, which
should approach to zero as $r\rightarrow \infty$. This is contradictory to the
original assumption of a finite saturation $B_{\infty}$. To be consistent,
then,
$B_{\infty}$ has to continuously vanish as $L\rightarrow \infty$.  Such a
conclusion holds for
a general spin state, no matter whether there is a real long-range order or
not. The energy $E$ lying in the regime $E_G^0>E\geq E_G$ should not
affect the decay significantly even when it is close to the ground state
energy, since the detailed spin fluctuations there are not
crucial. The actual decay of  $G_{1\sigma}$ could be even faster with the
increase of $|x_j-x_i|$ when all the paths are summed up in (\ref{epath})
which
contributes additional phase-frustration due to the strong-oscillation nature
of those  phase strings.

Therefore,  due to the phase string (\ref{estring}),  the propagator $G_{1
\sigma}$ has to  vanish at large
distance.  According to the discussion at beginning,  this in turn means either
$Z(E)=0$ at $E\rightarrow E_G$, or the quasiparticle spectrum
$E_k$ is dispersionless near the bottom.   Of course, the above discussions
cannot directly distinguish whether the true ground state of a one-hole doped
antiferromagnet is extended with $Z=0$ or localized with the dispersionless
$E_k$. In both cases, a ``bare'' hole loses
its coherence in the sense that it  cannot travel over a large distance near
the ground-state
energy. In the latter case, the quasiparticle hole would  have an
infinite effective mass, which could result in a divergent single-particle
density of states near the ground-state energy $E_G$. In this case,
finite-size
numerical calculations should give a reasonable account for the density of
states near the band bottom since $Z\neq 0$. However, a divergent density of
states or infinite effective mass is not supported by the numerical
calculations in the 2D case\cite{dagotto}. Furthermore, in 1D
the ground state has been already known to be extended from the exact
solution\cite{lieb}. Thus at dimensionality $\leq 2$, we conclude $Z=0$.

In our above demonstration, the conditions (\ref{e2}) and (\ref{e3})
are crucial in causing the unrepairable phase-string defects. They are related
to the intrinsic properties of the $t-J$ model. On the other hand, the property
of $|\psi_0>$ {\it as the ground-state of the undoped antiferromagnet}
actually
does not play a crucial role here.  Hence,  the whole argument
about $Z(E_G)=0$  should still remain robust at least  in the weakly-doped
case.  At finite doping,  additional
phase due to the fermionic-statistics among holes will appear in the
matrix (\ref{e3}),  and if the density of holes is dilute enough,  such a sign
problem should not weaken the afore-discussed phase-string effect.
On the other hand,  if  holes are dense enough in 2D,  it is not difficult to
find the strong mixing of the phase string with the statistic phase. In this
case,  new approach would be needed in order to clarify this issue.  But there
is no such a problem in 1D\cite{weng1}.

$Z=0$ means that there is no
overlap between the ``bare'' hole state $c_{i\sigma}|\psi_0>$ and the true
ground state. The doped hole will have to induce a {\it global}
adjustment of the spin background in order to reach the ground state. This
implies that by starting from $c_{i\sigma}|\psi_0>$,  one cannot get access
to  the ground state {\it perturbatively},  as there is no zeroth-order
overlapping. Thus, in order to correctly understand the long-wavelength,
low-energy physics of the weakly-doped $t-J$ model, a non-perturbative-minded
approach will become {\it necessary}.  Of course,  $Z=0$ itself does
not tell how the non-perturbative approach should be pursued.
One has to go back to the original source, i.e., the phase string introduced
by hole, which causes $Z=0$.  The nonlocal phase string (\ref{estring})
suggests that in the real ground state some nonlocal phase adjustment (i.e.,
phase shift) should appear.  In a separate paper\cite{weng2}, we shall show
that by eliminating the phase string (\ref{estring}) through
 some nonlocal transformation, a non-perturbative scheme can be obtained.
In this new scheme, the sign problem in 1D will be totally resolved, while
in 2D some residual topological phases, reflecting nonlocal coupling between
spin and charge degrees of  freedom, will  be determined.  We argue there
that this formalism provides a unique way to get  access to the low-lying
states of the $t-J$ model {\it perturbatively}.

In conclusion,  we have given a rigorous demonstration that for a one-hole
doping problem in the  $t-J$ model at general dimensionality, either the
spectral function $Z=0$ at $E=E_G$, or the quasiparticle is localized in
space with an infinite effective mass.  Such a  surprising result is caused by
a peculiar phase string induced
by the bare hole, which is revealed by explicitly tracking the Marshall sign
hidden in the spin background.  The present work implies that the doped $t-J$
model should not be pursued by a perturbative-minded approach in the usual
spin-hole representation,  and it points to the potential
perturbation-applicable scheme as the one with  the nonlocal phase string
being eliminated.

\acknowledgments
The present work is supported partially by
Texas Advanced Research Program
under Grant No. 3652182, and
by Texas Centre for Superconductivity at University of Houston.\\

* Permanent address: Department of Physics, University of Science and
Technology
of China, Hefei, Anhui 230026, China.


\begin{references}
\bibitem{remark} Here one should be cautious
about some special cases where $\rho(E_G)=0$.  For example, it occurs at a
dimensionality higher than 2D for a parabolic spectrum, and in 2D for a
massless Dirac spectrum $E_k$. But in those cases,  $G_{1\sigma}/\rho(E)$
would be extended as $E\rightarrow E_G$ if $Z_k\neq 0$, and a
strong energy dependence of the decay rate should be still revealed in
$G_{1\sigma} $ at large $|x_j-x_i|$ when $E\sim E_G$.
\end{references}
\end{document}